\documentclass{emulateapj}

\usepackage{epstopdf}

\def\arcsec{$^{\prime\prime}$}

\shorttitle{Proper Motions and Dynamics in M71}
\shortauthors{Samra et al.}

\begin{document}

\title{Proper Motions and Internal Dynamics in the Core of the Globular Cluster M71}

\author{Raminder Singh Samra\altaffilmark{1}, Harvey B. Richer\altaffilmark{1}, Jeremy S. Heyl\altaffilmark{1}, Ryan Goldsbury\altaffilmark{1}, Karun Thanjavur\altaffilmark{2}, Gordon Walker\altaffilmark{1}, and Kristin A. Woodley\altaffilmark{1}}

\altaffiltext{1}{Department of Physics and Astronomy, University of British Columbia, Vancouver BC V6T 1Z1 Canada: rsamra@phas.ubc.ca, richer@astro.ubc.ca, rgolds@phas.ubc.ca, heyl@phas.ubc.ca,  gordonwa@uvic.ca, kwoodley@phas.ubc.ca}
\altaffiltext{2}{Canada-France-Hawaii Corporation,
Waimea, HI 96743-8432, United States: karun@cfht.hawaii.edu}

\begin{abstract}
We have used Gemini North together with the NIRI-ALTAIR adaptive optics imager in the H and K bands to explore the core of the Galactic globular cluster M71 (NGC 6838). We obtained proper motions for 217 stars and have resolved its internal proper motion dispersion.  Using a 3.8 year baseline, the proper motion dispersion in the core is found to be $179\pm17$ $\mu$arcsec/yr.  We find no evidence of anisotropy in the motions and no radial variation in the proper motions with respect to distance from the cluster center. We also set an upper limit on any central black hole to be $\sim$150 M$_{\odot} $ at 90\% confidence level.  
\end{abstract}


\section{Introduction}

It has been recently shown that the cores of globular clusters (GCs) may be hosts of Intermediate Mass Black Holes (IMBHs), which have masses from $\sim$100 M$_{\odot} $ to several 10$^{4} $ M$_{\odot}$ \citep[e.g. ][]{noyola08, NGC6388}. If they are present, they could provide the missing link between supermassive black holes (SMBHs) in the cores of galaxies to stellar mass black holes \citep[e.g. ][]{maccarone07}.

IMBHs can be detected in several ways; a rise in the stellar density towards the cluster centre \citep{lbata}, x-ray or radio emission \citep{ulvestad07} or via a kinematic signature. Most GCs contain little gas therefore an IMBH would likely be most easily detected in the kinematics of its stars. Such signatures could be stars moving faster than the escape velocity of the cluster, a steep increase in the velocity dispersion profile towards the center, or observing orbital motion about the central BH, as seen in the center of the Galaxy \citep{ghez05}.   N-body simulations   \citep{baumgardt05,trenti07}  have suggested that clusters with large cores potentially harbour IMBHs, which provide an energy source to \lq fluff-up\rq ~the central region of the cluster.  Within this context,  M71's core radius is $\sim$38,\arcsec\ \citep{harris10} larger than $\sim$70\% of all Galactic GCs.  


\begin{figure}
\centerline{\includegraphics[width=8cm,clip,trim=70 0 0 0]{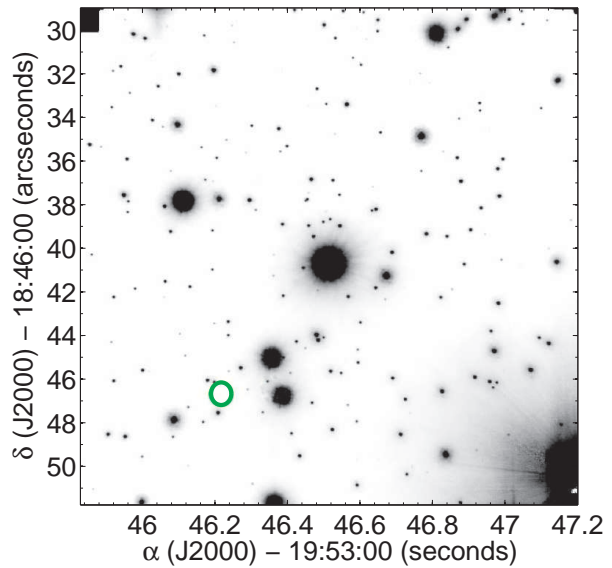}}
\caption{The core of M71 as imaged by the Gemini Telescope with the NIRI/ALTAIR AO system in $H$ band. The telescope's dither pattern resulted in a field that is 23.5 arcseconds on each side.  The center of the small circle indicates the cluster center from \cite{goldsbury10} with a $\sim$0.5\arcsec\ uncertainty.}
\label{fig:NIRI}
\end{figure}


An existing proper motion (hereafter PM) study was done by \cite{cudworth85} where it was determined that M71 exhibits a low space velocity for a GC and that it was on a disk-like orbit in the Galaxy. \cite{cudworth85} also calculated the PM dispersion for stars 100\arcsec\ from the center and found it to be 180 $\pm$ 30 $\mu$arcsec/yr.

Modern values of the parameters describing M71 can be found in \cite{harris10}. The distance to M71 was determined several times, \cite{hartwick71} place the cluster at 4100 $\pm$ 400pc, \cite{geffert00} place it at 3600 $\pm$ 250pc.   The most recent and accurate center of M71 was found by \cite{goldsbury10} who placed the center at $\alpha$= 19h53m46.25s and $\delta$= +$18^{o}46' 46.7$\arcsec  (J2000) with an error of $\sim$0.5\arcsec.

\section{Data Reduction and Analysis}
In 2005, we observed M71 with the Near Infrared Imager (NIRI) in conjunction with the ALTAIR adaptive optics system on the Frederick C. Gillett Gemini North Telescope on Mauna Kea (see Figure \ref{fig:NIRI}). The observations were carried out in the $H$ (1.65$\mu$m) and $K$ (2.2$\mu$m) bands in the core of M71 centered on the star AH 1--83 \citep{hartwick71}. NIRI was set to f/32 to image a field of view of 22\arcsec\ x 22\arcsec\ with a pixel scale of 0.0219\arcsec/pixel and a twelve step dither pattern was employed with spacings separated by 1\arcsec\ in both RA and Dec.  Queue mode observations ensured that for each of our imaging nights, the typical seeing was the best possible at Gemini, $\leq$ 0.6\arcsec\ in R band.  Table \ref{tab:logs} contains the observational logs from the three different imaging epochs, all of which used the same instrumental settings aside from the total integration time and filter.


\begin{deluxetable}{llll} 
\tablecolumns{4}
\tabletypesize{\small}  \tablecaption{Observational Logs\label{tab:logs}}
\tablewidth{0pt}
\tablehead{
 \colhead{Date}& \colhead{Filter}  & \colhead{Exposure (ksec)} & \colhead{FWHM (\arcsec)} \\
}
\startdata
August 2005 &   H   & 15    & 0.085   \\
August 2005 &   K   &  16   &   0.081   \\
June 2007    & H     &14     & 0.083\\
June 2009    & K     & 23 & 0.077 \\
\enddata
\end{deluxetable}

\begin{figure}
\hspace{-2.0em}\includegraphics[height=5.5cm,width=5.6cm,clip,trim=30 3 0 30]{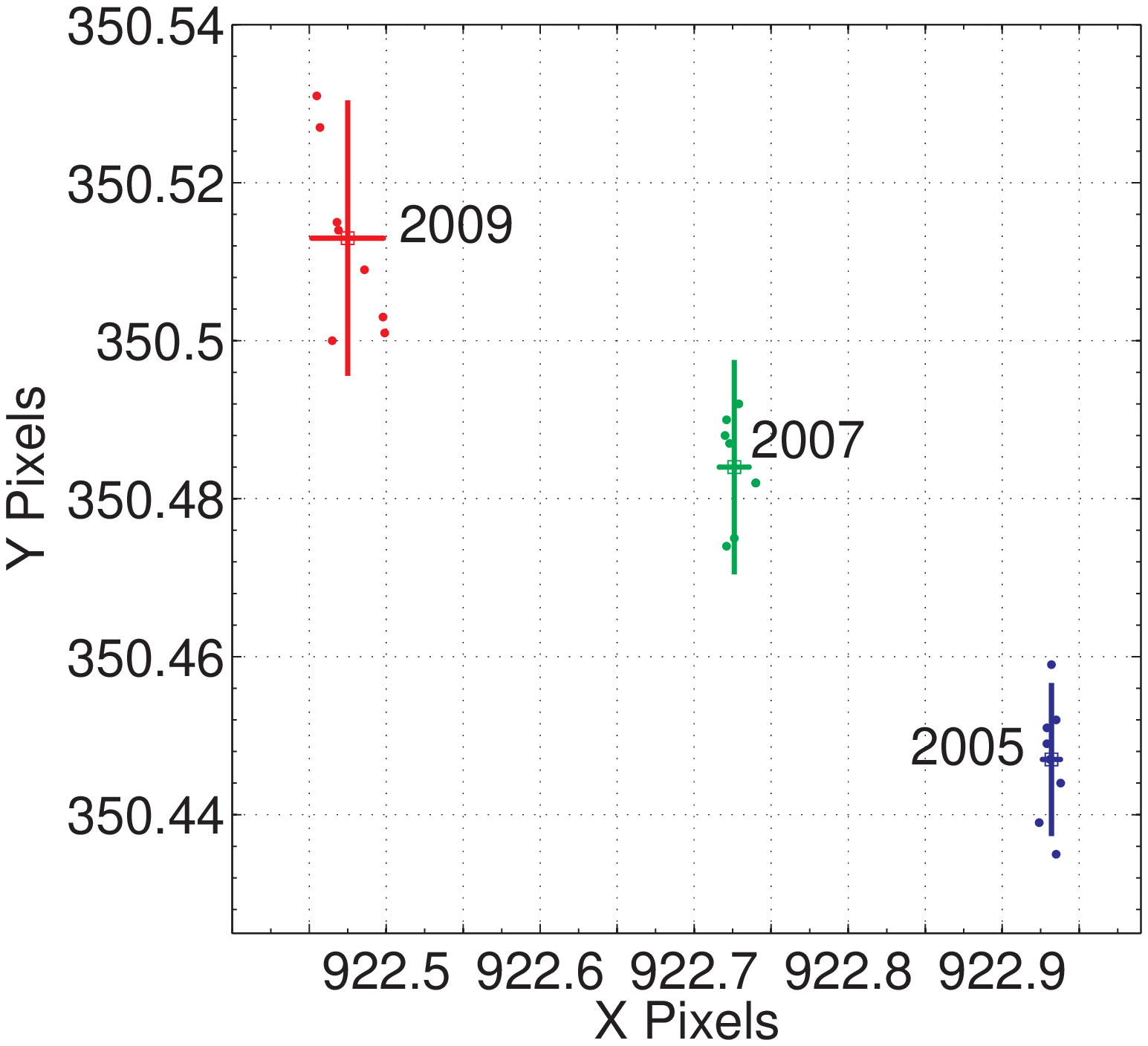}\includegraphics[height=5.5cm,width=4.1cm]{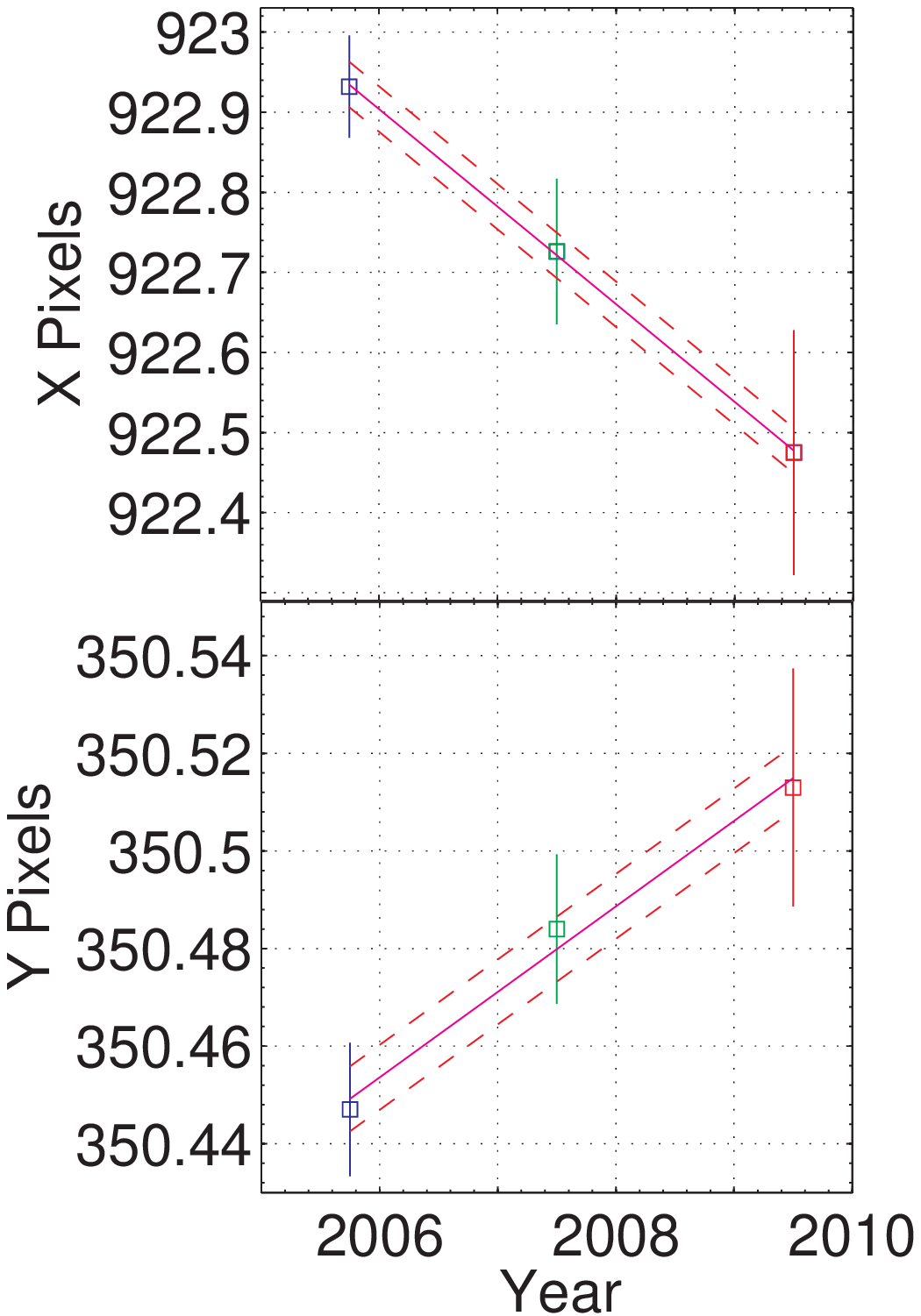}
\caption{ \textit{Left:}  The relative location of one cluster member across the field in each of the three imaging epochs. The colored dots represent the derived locations from each ALLFRAME run. The error in the final position is determined from the spread in the points for each epoch. \textit{Right:} The X and Y positions of the same star as a function of time. The PM was found from fitting a straight line through those points, the slope of the line multiplied by the NIRI f/32 plate scale gives the PM. The dashed lines represent one-sigma errors.   }
\centerline{}
\label{fig:threepoint}
\end{figure}

The data were reduced using the GEMINI IRAF package which flat fielded and dark subtracted the raw NIRI data.  We used the DAOPHOT II \citep{stetson} software package to find stars and perform aperture photometry on the individual frames.  We then used the DAOMATCH/MASTER tasks to locate the same stars across each of the individual frames, and iteratively rejected stars that were not located on each frame within 1~pixel.  Once we had all of the stars on the same coordinate system and the shifts between each of the images due to the telescope\rq s dither pattern were calculated, a median combined image of the cluster was constructed.  This median image was then used to create a deep finding list of 217 stellar-like objects in the NIRI field.  ALLFRAME was then run and returned the coordinates for each star in the coordinate system of the median combined image; this package was then used for each of the other imaging epochs to find the coordinates of each stellar object.  The error in each star's position for each epoch was obtained by running ALLFRAME in sets of $\sim$20 frames for each individual epoch. This resulted in 7-8 positions for each star for each of the three epochs; the standard deviation of the positions was the resultant error in the position. PMs were then derived by fitting linear time dependent curves for each star's position as shown in Figure \ref{fig:threepoint}.

\begin{figure}[h]
\hspace{-0.93em}\centerline{\includegraphics[width=10cm,clip,trim=0 0 0 0 0.25in]{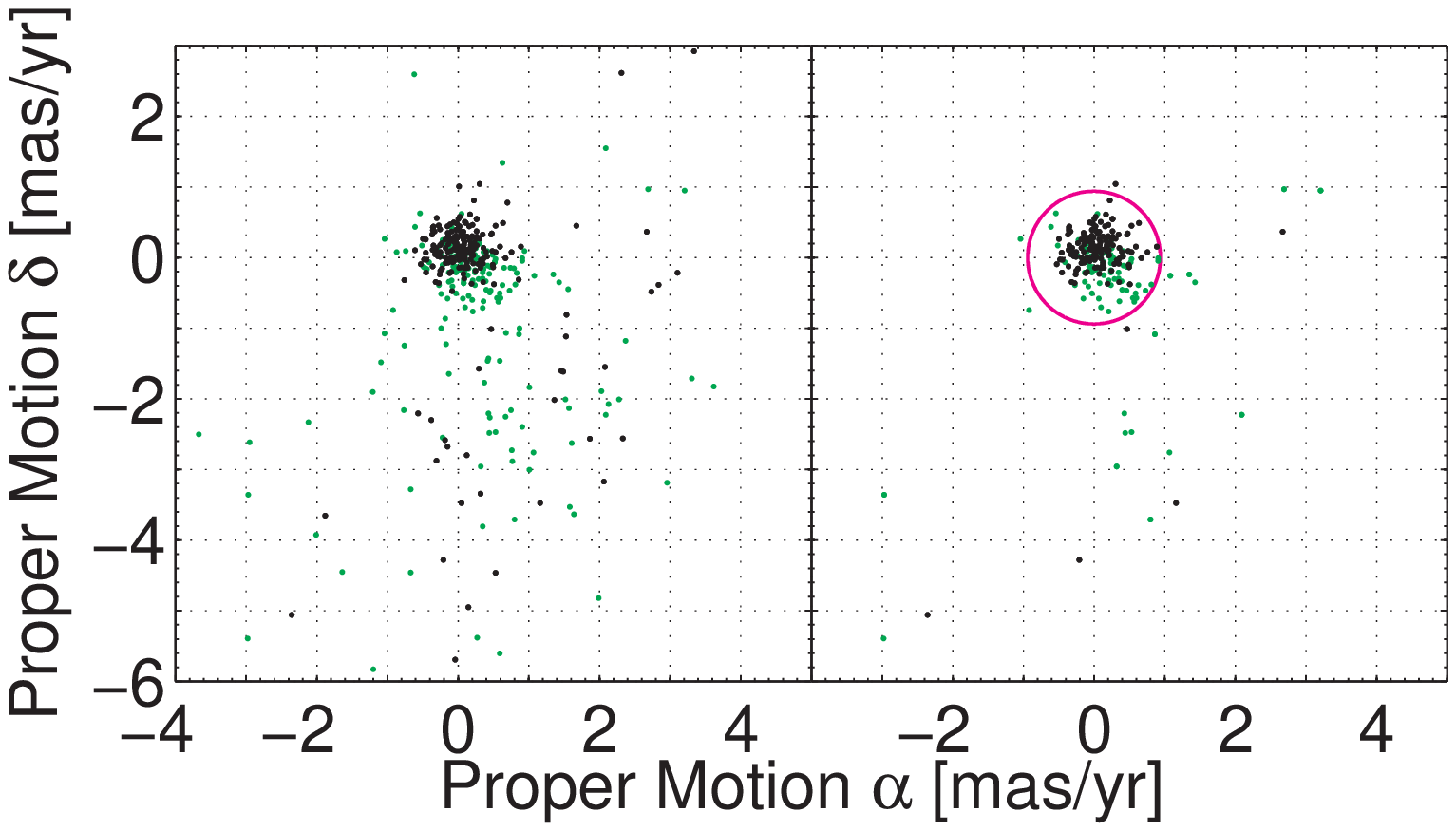}}
\centerline{  \hspace{-1.7em}\includegraphics[height=12cm,clip,trim=3cm 0 0in 0.00in]{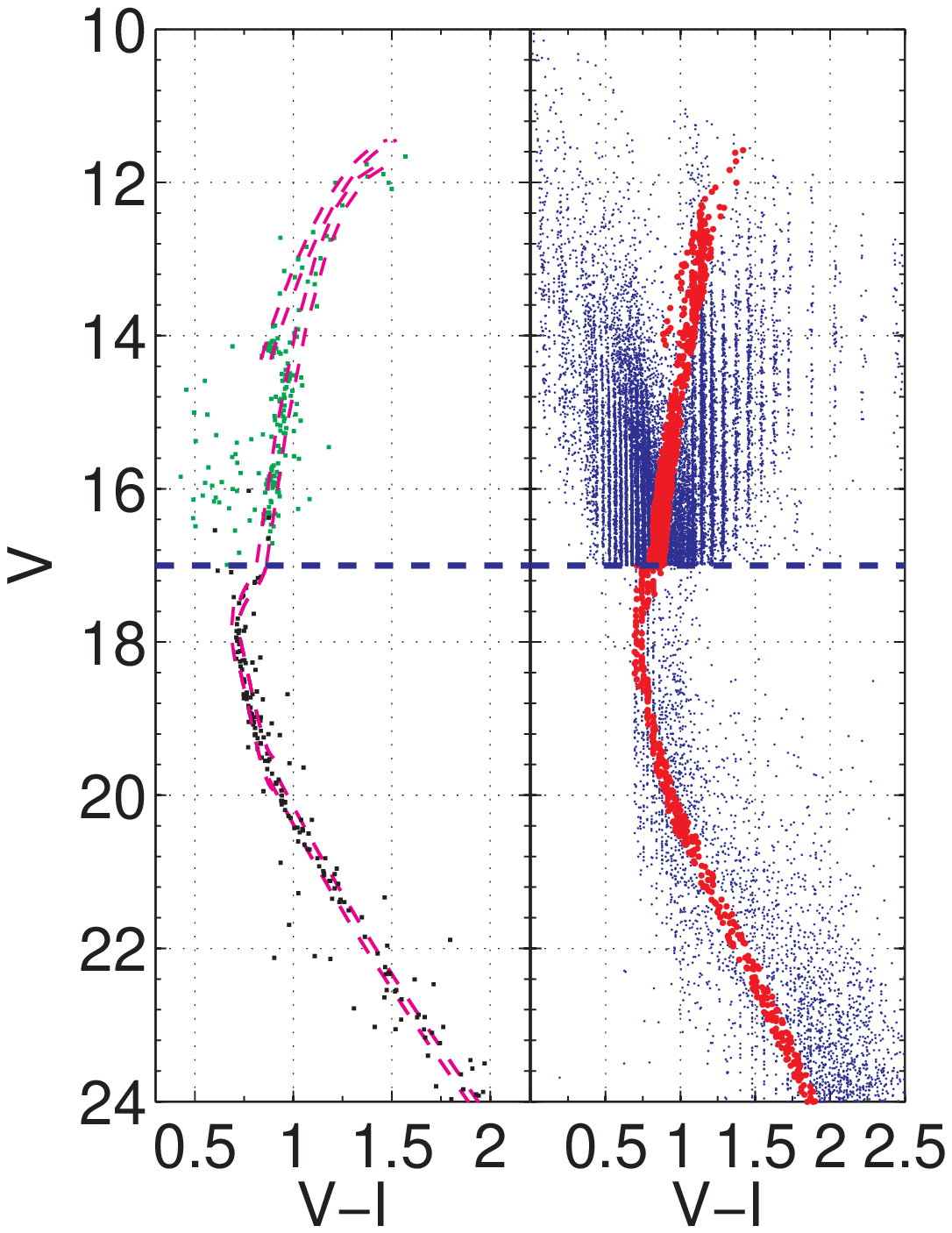}}
\caption{
{\it Top Left:} The differential PMs relative to the cluster mean in RA and Dec from the NIRI field (black dots) and all the stars in the field of \cite{cudworth85} for which we have ACS photometry (green dots).
{\it Bottom Left:} The CMDs from ACS photometry of both our field and \cite{cudworth85}.  The horizontal blue line represents the cut-off of both studies. We observed stars with V \textgreater ~17 and \cite{cudworth85} only had stars with V \textless~ 17. The curves represent a fiducial MS, likely cluster members are located along this curve.
{\it Top Right:} The PMs of those stars that satisfy the color criterion depicted in the bottom-left panel.  The circle depicts an estimate of the escape PM  from \cite{gnedin02}. {\it Bottom Right:} The CMD of the Besan{\c c}on models ({\S} \ref{sec:HVS}) for two simulations of the NIRI and Cudworth's field.}
\label{fig:cmds}
\end{figure}

We then used online data from the ACS Survey of Galactic Globular Clusters in the HST filters F606W and F814W transformed to V and I \citep{sarajedini07, anderson08} to construct a color magnitude diagram (CMD) of our field. The CMD of the HST stars contained in the NIRI field is shown in Figure \ref{fig:cmds}. 
We used stars which had total PMs less than 0.8~mas/yr (half the PM dispersion of all stars) to define a fiducial main sequence (MS).   Our cluster sample consists of all stars within 0.05 magnitudes in color of the MS fiducial, regardless of their measured PM.

A previous PM  study of M71 was performed by \cite{cudworth85}, we used the 358 stars listed in that paper's table to verify our PMs. We first matched Cudworth's stars to the ACS photometry to obtain colors for 208 stars. We then performed a color-cut for stars within 0.05 magnitudes of the RGB to define a PM winnowed sample.  
We then removed all stars from both the NIRI and Cudworth's  winnowed samples which had PM errors greater than 2$\sigma$. This resulted in 150 cluster members for our NIRI field and 94 stars from Cudworth's sample. These 94 stars ranged in radial distance between 6\arcsec-150\arcsec ~from the cluster center, 5 of which were in our NIRI field for which we did not have PMs due to them being saturated on our frames. Figure \ref{fig:cmds} has both our NIRI field PMs and Cudworth's PMs included, the final PM sample we used for this study are the stars in the top-right figure. 


\section{Results}
\subsection{Proper Motion Distribution}

To quantify the velocity dispersion of the cluster, one is tempted to use the standard deviation.  However, a single high-velocity interloping star in the sample can skew the standard deviation to arbitrarily large values.  Therefore we seek an estimator of the dispersion that is robust to outliers.   
 \citet{rousseeuw93} propose $Q_{n}$, motivated by two of its strengths --- its ability to deal with skewness and its efficiency with gaussian distributions.  It is defined over all the pairs of stars in the sample, 
\begin{equation}
Q_{n} = a \times \textrm{first quartile of} ~(|\mu_{i} - \mu_{j}|) : i < j )
\label{eqn:qn}
\end{equation}
where $a$ is a constant dependent upon the size of the sample (for large samples $a\rightarrow2.2219$) and $\mu_{i} $ and $\mu_{j}$ are observed PMs in a particular direction of pairs of stars in our sample.   One further advantage of the $Q_{n} $ estimator is that it approximates the standard deviation for distributions which are significantly different from normally distributed distributions such as those close to a uniform distribution.  We treat $Q_{n} $ as a proxy for the standard deviation for the remaining analysis.  For example, we correct for the errors in the observed PMs from the estimated errors to arrive at the PM dispersion: 
\begin{equation} \label{eqn:qndisp}
Q_{n}^{2} =  Q_{n_{o}}^{2} - \frac{1}{N}\sum \epsilon_{\mu_i}^{2}
\end{equation}\
where $Q_n $ is the error corrected dispersion, $Q_{n_{o}} $ is the observed non-error-corrected PM dispersion and $\epsilon_{\mu_i} $ is the error associated with the PM of each star. The error in the dispersion was calculated by bootstrapping the PMs and their errors 100,000 times and taking the standard deviation of the bootstrapped dispersions. This results in a one-component core dispersion of $Q_{n}$ = 179 $\pm$ 17 $\mu$as/yr. Using the familiar standard deviation method to calculate the dispersion we obtain a value of $\sigma$ = 185 $\pm$ 18 $\mu$as/yr, that is after removing stars with PMs greater than the cluster's escape PM.  


We used Cudworth's stars (see the top right panel of  Figure \ref{fig:cmds}) to determine the PM dispersion for stars which had radial distances greater than our NIRI field and were likely cluster members. We found for stars in the inner 70\arcsec ~the dispersion was 184 $\pm$ 37$~\mu$as/yr, for stars with distances greater than 70\arcsec ~and up to 150\arcsec ~the dispersion was 281 $\pm$ 33~$\mu$as/yr. Results which are similar to the published values from \cite{cudworth85}.   

At the distance of 4 kpc our PM corresponds to a velocity dispersion of  3.3 $\pm$ 0.4 km/s, a value greater than the radial velocity dispersion of 2.0 km/s obtained by \cite{rastor91} and 2.8 $\pm $ 0.6 km/s from \cite{peterson86}. However those velocity dispersions were calculated for giant stars which were far away from the center of the cluster; a constant dispersion is still within the uncertainties. 

 \begin{figure*}
\centerline{ \hspace{-0.0013em} \includegraphics[height=6cm,width=6cm]{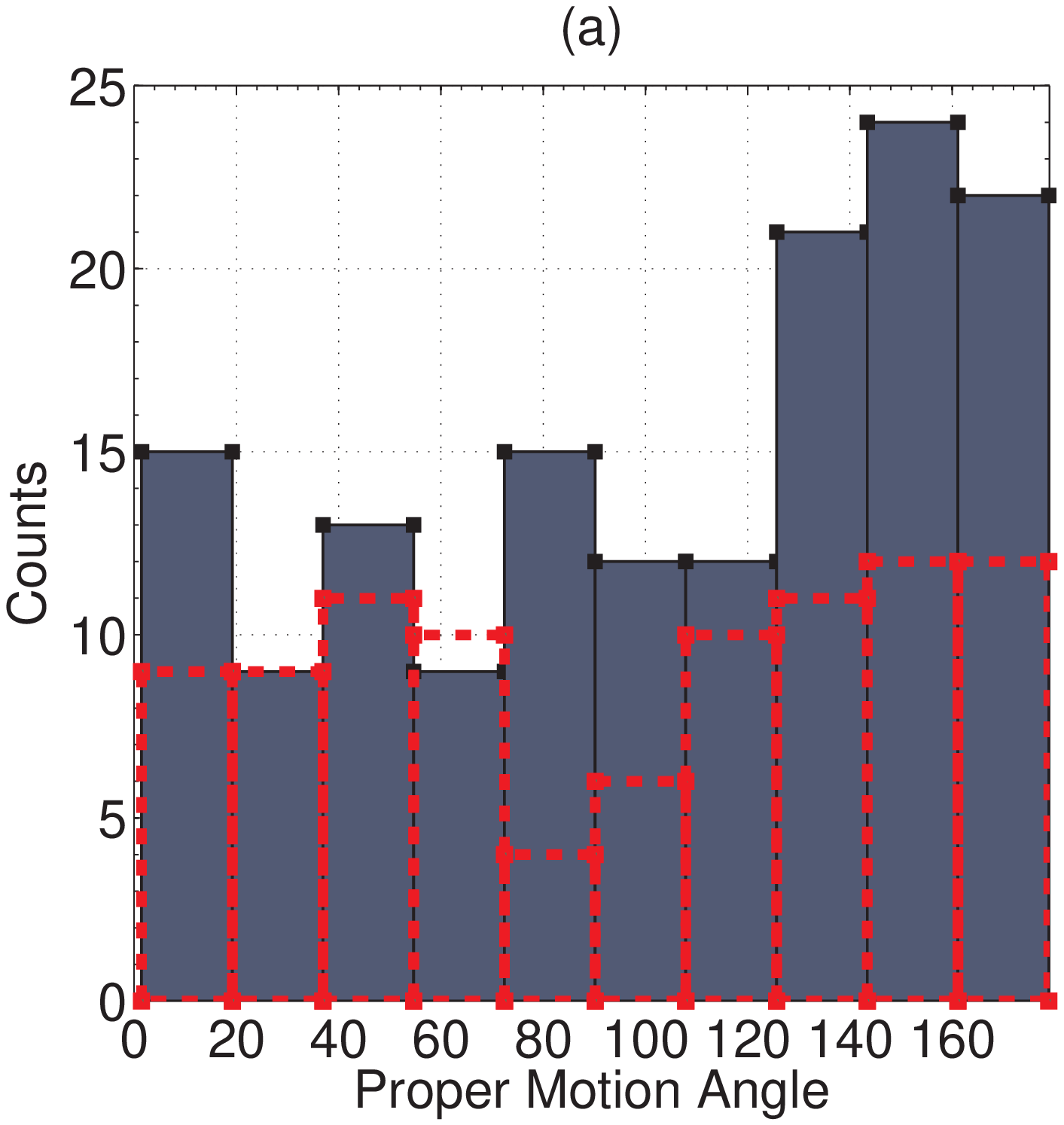} \includegraphics[height=6cm,width=6cm]{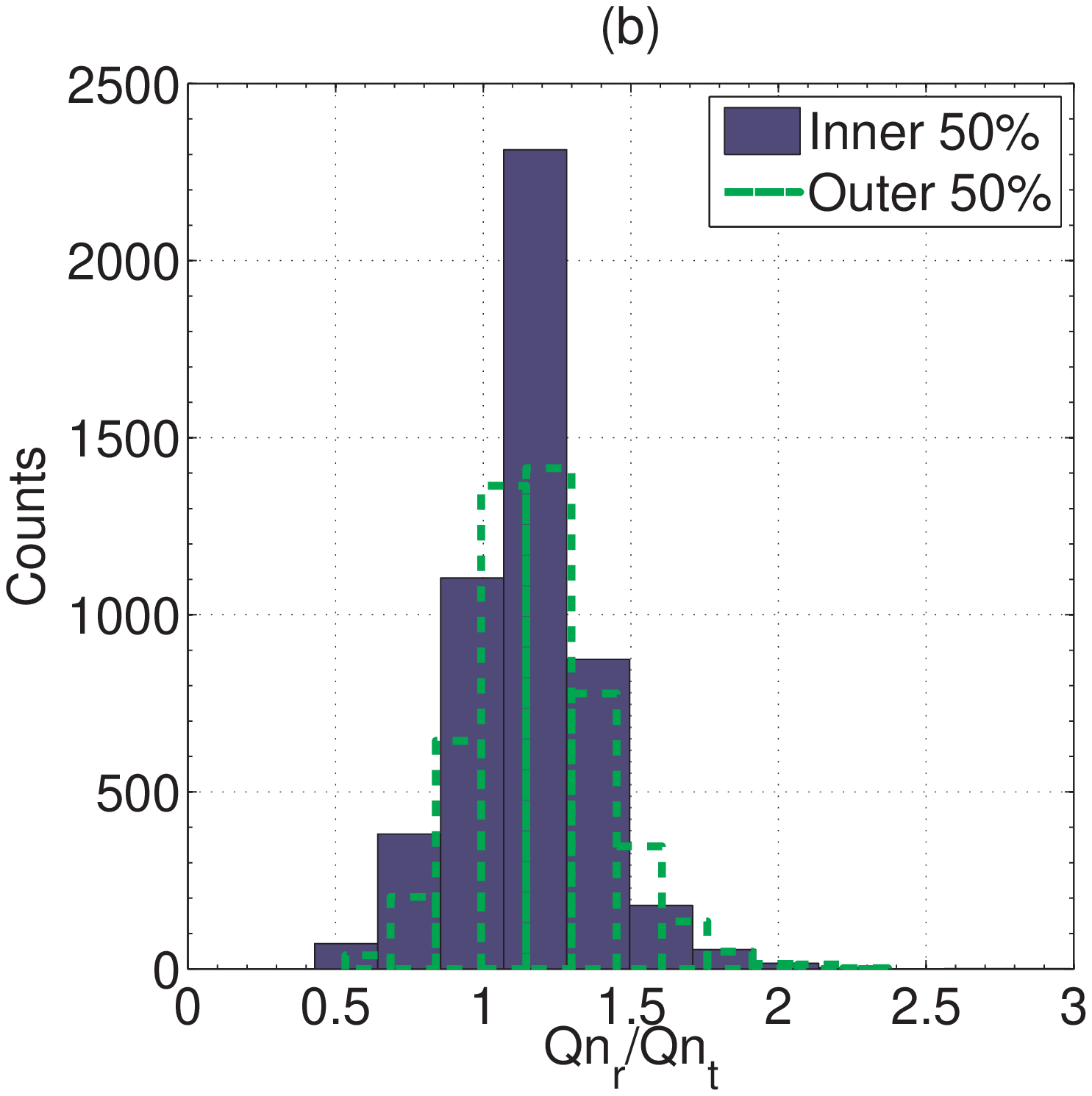} }
\centerline{\hspace{-1em} \includegraphics[height=6.5cm,width=6cm,clip,trim=0 0.0 0 0.0]{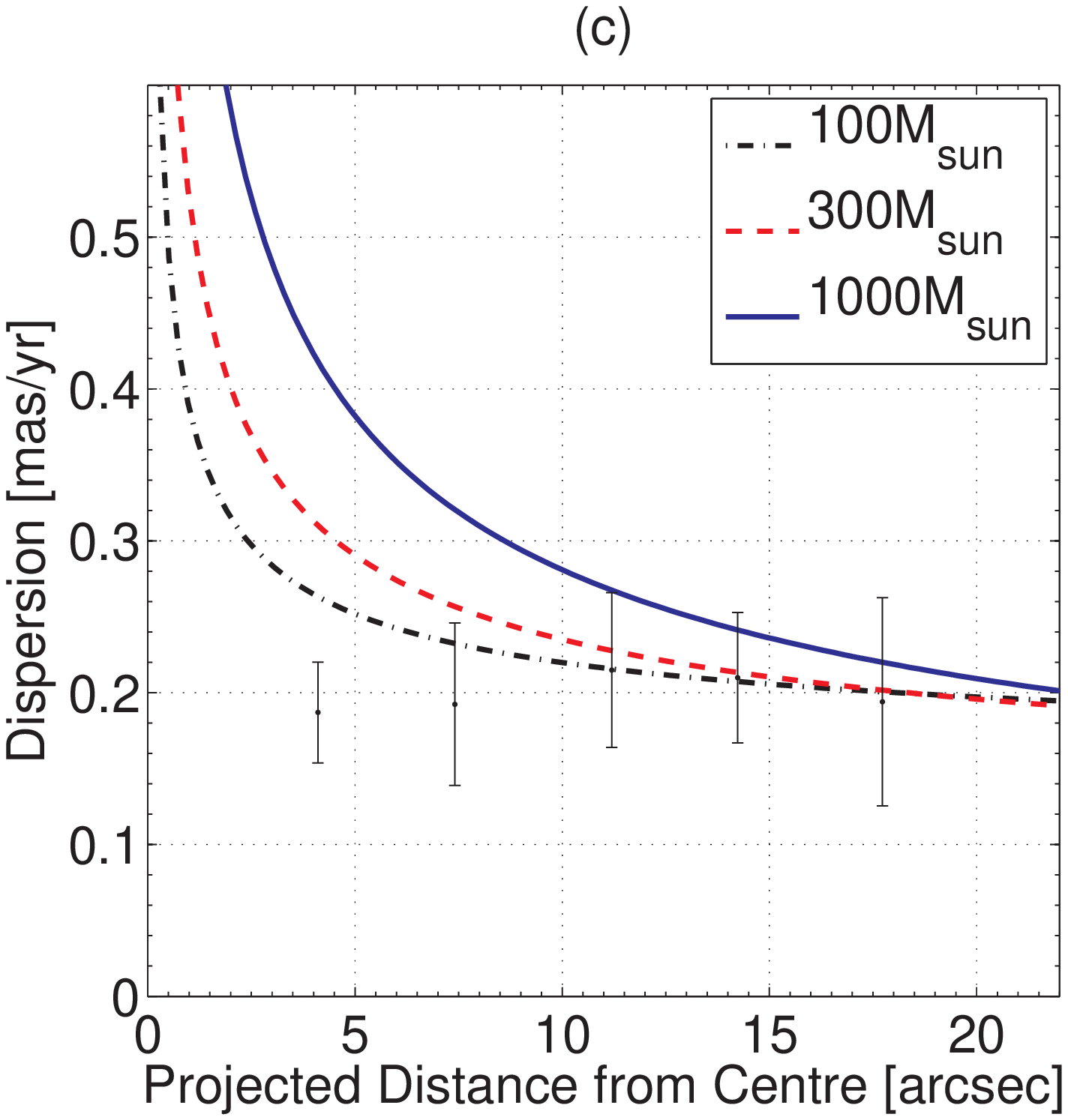}
\includegraphics[height=6.5cm,width=6.2cm,clip,trim=0 0 0 0]{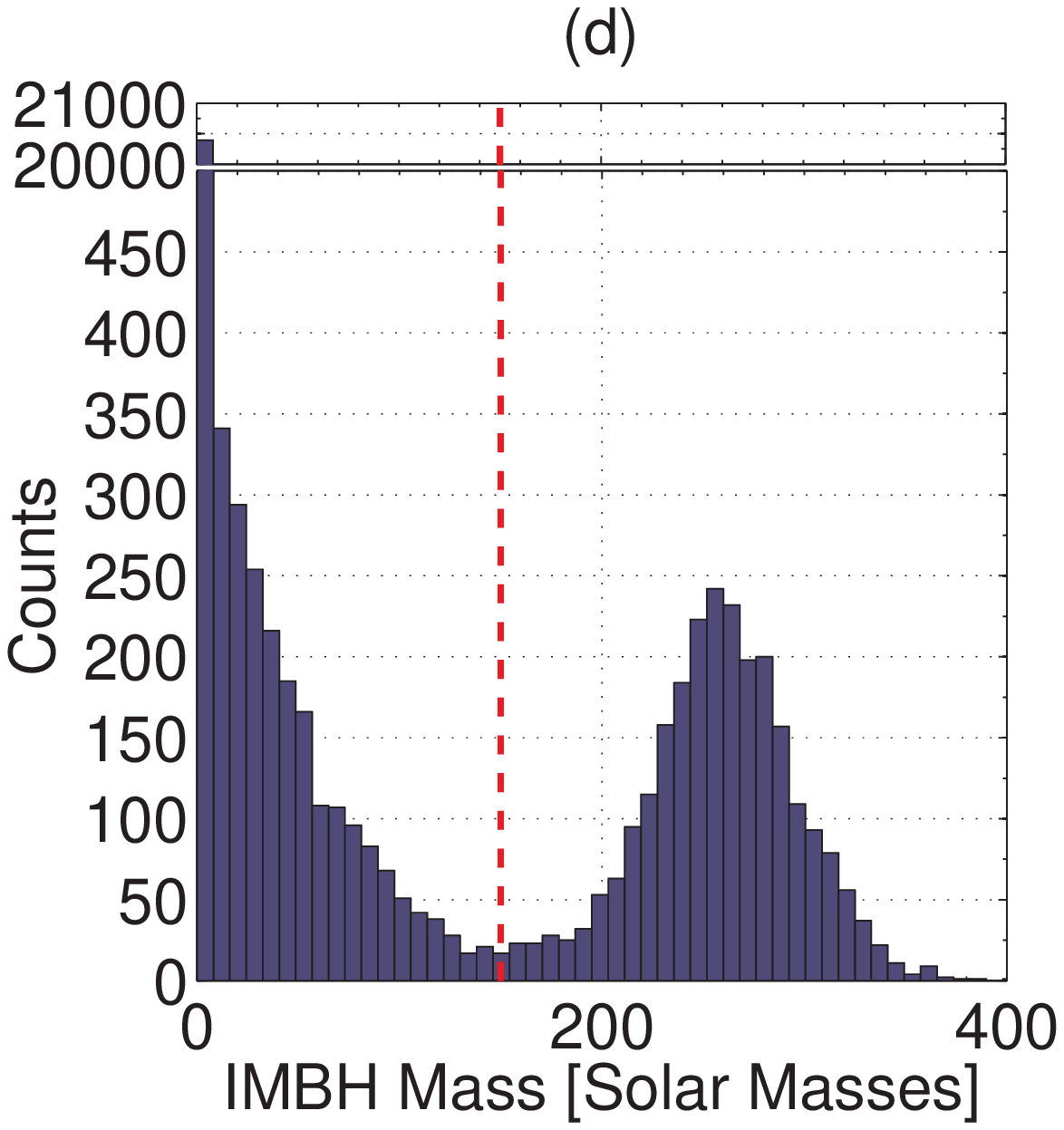}
\includegraphics[height=6.5cm,width=6.2cm,clip,trim = 0 0 0 0 ]{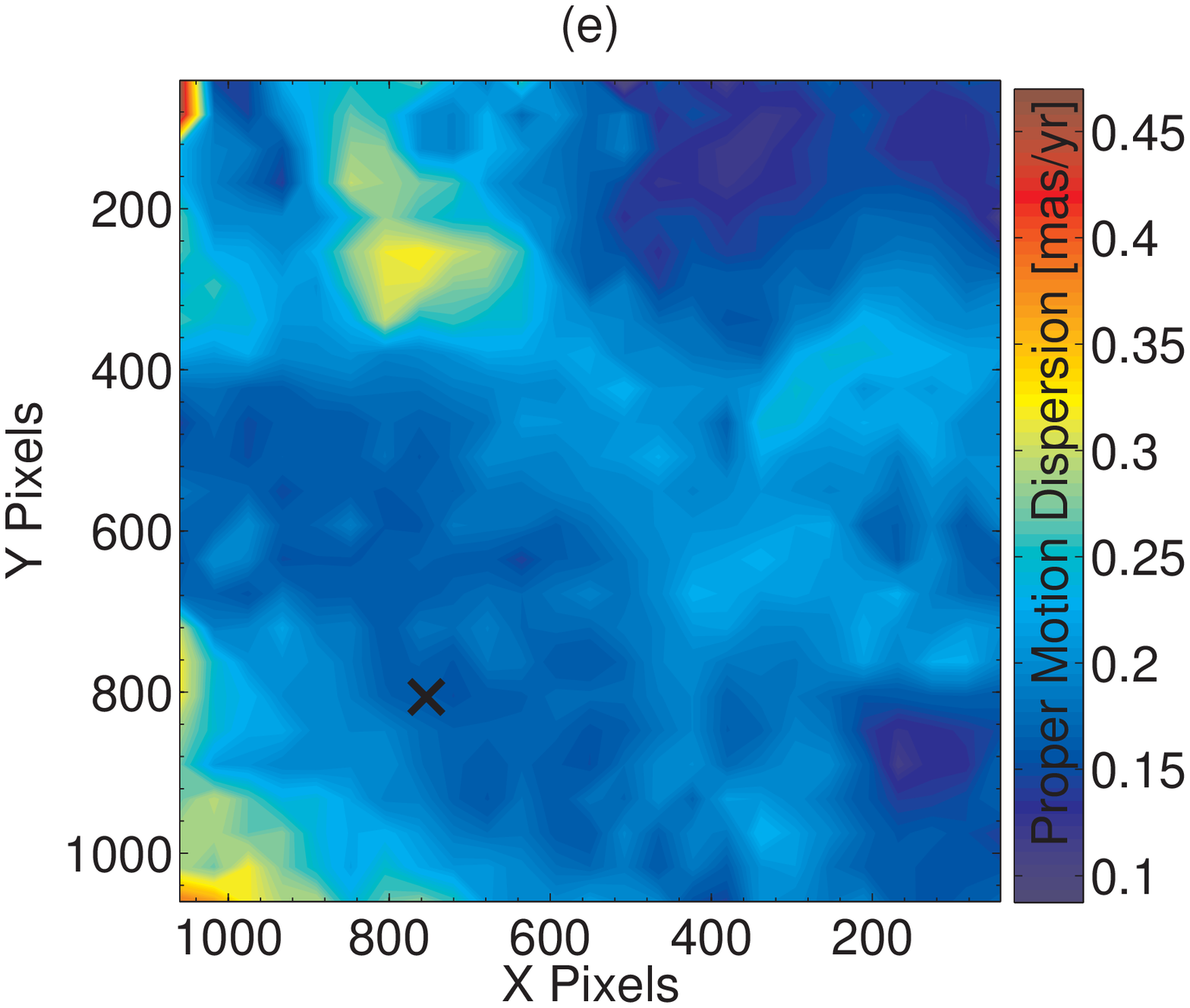}}
\caption{\textit{a: }Histogram of the PM angle with respect to the center of the cluster. Blue histograms are for stars in the NIRI field, the dashed red line histogram is for stars from \cite{cudworth85}.  \textit{b}:Histograms for the values $Q_{n_r}/Q_{n_t}$ resulting from the bootstrap resample of the data. The radial and tangential values of $Q_{n} $ are consistent with  unity within 1$\sigma$ for all stars.  \textit{c: } The observed PM dispersion and three different IMBH models. \textit{d: }  The results from bootstrapping the proper motion data 25,000 times, the IMBH mass from each fit is binned here. The leftmost bin from 0 $\leq M_\odot \leq 5 $ peaks at 20,500 counts, implying that a very small central IMBH is the preferred fit.  The red line represents the 90$^\textrm{{th}} $ percentile at 150M$_{\odot}$.  \textit{e:} A dispersion map of the field, obtained by running a circle with a 4\arcsec\ diameter across the field and calculating the dispersion in each area. Observationally, there are no signs of radial dependence or an increase in dispersion towards the center which is represented by the X. } 
\label{fig:isotropy}
\end{figure*}

\subsection{High Proper Motion Stars}
\label{sec:HVS}

From the CMD in the bottom left panel of Figure \ref{fig:cmds}, we find 7 stars which have high PMs and survived the winnowing and error cuts. If any of these stars are escaping the cluster, we could determine the cluster's evaporation timescale purely from  observations. However our observed field lies at low Galactic latitude ($b = -4.56$ degrees), therefore the field may be subject to heavy contamination from the Galactic field as noted by \cite{cudworth85}. To check for possible interlopers, we used the Besan{\c c}on models of stellar population synthesis of the Galaxy from \cite{robin03}. The models depend on the Galactic coordinates, field size, photometric colors, and the extinction. The value for extinction in the direction of M71 was found to be 0.19~mag/kpc which was calculated using the reddening constant of $3.14\pm0.10$ from \cite{schultz75}, the distance, and the reddening from \cite{harris10}.  For a field 100 times the area of our NIRI field, the model yields about 5800 stars, 625 of which lie on our cluster's main sequence. Therefore we can expect $6-7$ of the stars in our NIRI field to be interloping field stars (bottom right panel of Figure \ref{fig:cmds}). It is therefore difficult to conclude that we have found any escaping stars. 

We also ran a simulation for a field 150 times the size of the ACS field; we find $\sim$1900 stars in this simulation which lie on the  winnowed CMD for  Cudworth's stars. Scaling this down to the ACS field we expect to find $\sim$13 interloping stars.  From the PM distribution we find 16 potential escapers, it is once again difficult to conclude we have observed any escaping stars.

\subsection{Proper Motion Isotropy}
Using the PM data, we determine the PM vector of the individual stars in the cluster. We calculated the PM angle with respect to the cluster center for three groups of stars; half of the NIRI field contained stars within 11\arcsec ~the other half were from 11-22\arcsec, we also calculated the angles for stars from \cite{cudworth85} which we determined to be cluster members.  We found that stars in the three different radial distances had PM angles which were consistent with isotropy. 
We performed a Kuiper test on the data from Figure  \ref{fig:isotropy}a, which is a test similar to the Kolmogorov-Smirnov test but more sensitive to differences in the wings of the distributions. This test is better suited for distributions which are cyclic in nature as in our case. The $p-$value, comparing a flat distribution, is 0.17, for stars in the inner 11\arcsec ~and p=0.21 for stars in the outer 11\arcsec; finding no significant difference when compared to a flat isotropic distribution. We have a similar result when we extend our field from the inner 22\arcsec ~up to 150\arcsec, we then have a p-value of 0.14. Figure \ref{fig:isotropy}a shows the distributions of PM angles for the NIRI field and from Cudworth.

As we transform coordinates between the multiple epochs, we use cluster stars as the reference points; therefore, any mean motion of the cluster is lost.  To characterize the anisotropy of the PMs, we measure the width of the PM distribution in the radial and tangential directions using $Q_n$.   The measured ratio of the radial to tangential widths $Q_{n_r}/Q_{n_t}$ is unity for an isotropic distribution.
We find the ratio of the two components $Q_{n_r}/Q_{n_t}$
  for three radial distances. For the inner 50\% we have $Q_{n_r}/Q_{n_t}$ = 1.15$\pm$0.22 and $Q_{n_r}/Q_{n_t}$ = 1.17$\pm$0.23 for stars for the outer 50\% of the NIRI field.  If we extend  our field up to $\sim$150\arcsec ~we have  $Q_{n_r}/Q_{n_t}$ = 1.03$\pm$0.13. The errors were obtained by bootstrapping our sample of measured PMs, and accounting for errors in the cluster's center and PMs.  Although the results for stars in our NIRI field  are above unity and may indicate a stronger radial component in the PMs, both results are still within 1$\sigma $ of unity. Figure \ref{fig:isotropy}b shows the distributions of the ratios of the radial and tangential dispersions from our NIRI field.

\subsection{An Upper Limit to the Mass of any Central IMBH}\label{sec:imbh}

IMBHs have been suggested to exist in several GCs such as G1 \citep{gebhardt, ulvestad07}, NGC 6388 \citep{NGC6388}, and Omega Cen \citep{noyola08},  however some of the claims are still disputed, notably the claim in Omega Cen \citep{anderson2010}. \cite{SAFONOVA10} extrapolate the SMBH correlations down to GC masses, determining a linear relation in the velocity dispersion and central black hole mass.  Using their relation and our observed velocity dispersion, we obtain an upper limit of a central black hole mass of 25 M$_{\odot} $ for M71.  It is risky, however, to extrapolate correlations between SMBHs down to IMBHs as their formation scenarios might be significantly different.   To determine if we have observed any evidence for a central IMBH in the core of M71, we divide our PM data for stars in the inner 22\arcsec ~into 5 radial bins. Each bin contains 30 stars, we then calculate the observed PM dispersion (Equation  \ref{eqn:qndisp}) and estimate the errors by bootstrapping and calculating the spread in the values of $Q_n$ (see Figure \ref{fig:isotropy}c). The observed PM dispersion is nearly constant throughout the inner 22\arcsec.

We have found that the velocity dispersion calculated by 
\cite{baumgardt04} in clusters with a black hole at the center can be modeled simply
as a function of the cluster half-light radius for stars which are close to the cluster center.  The resulting model contains an $r^{-1/2} $ Keplerian contribution from the black hole and a constant  dispersion from the stars only. To quantify our results we bootstrap the observed PMs and fit them to our model by minimizing the $\chi^{2} $ values and accounting for our error in the cluster center.  
  Using these bootstrapped data we find 
with 90\% confidence, that any black hole at the center of M71 must be less massive than $150 d_4^3$  M$_{\odot} $.   Where $d_4$ is the distance to the cluster divided by 4 kpc. From the histogram in Figure~\ref{fig:isotropy}d the most common result is the one with a very small black hole, where the PM distribution is constant with radius. Near the 90$^\textrm{th} $ percentile the value of $\sigma $ from our model vanishes and the observed PMs obey a Keplerian $r^{-1/2} $ distribution. 

One of the risks with binning data is that it is possible to over- or under-bin and mask or amplify any underlying features or noise. We approach this with two methods. First we run a circular bin across our field and calculate the dispersion in each bin to produce a dispersion map. Figure  \ref{fig:isotropy}e shows this map and it is clear the changes in dispersion are random fluctuations. 

Second, we calculate the average distance from the center,
\begin{equation}
\bar{r} =  \Sigma r / N
\end{equation}
and compare it to the proper motion weighted distance from the center,
\begin{equation}
\bar{r}_{|\mu|} = \frac{\Sigma r|\mu| } {\Sigma|\mu| }  .
\end{equation}
If the PMs are constant as a function of distance from the center, the two values should be equal.  However, if PMs are decreasing with increasing radial distance, larger radii get less weight so the value is less. We obtain $\bar{r}  = 11.81$\arcsec$\pm$  0.47\arcsec\ and $\bar{r}_{|\mu|}  = 11.50$\arcsec$\pm$ 0.83\arcsec.  The PM weighted distance is lower, indicative of higher proper motions at smaller radii, but a radially constant velocity dispersion is completely consistent within the uncertainties.

\section{Conclusions}

Using NIRI on the Gemini North Telescope, we have been able to resolve the internal PMs in M71. With a 3.8 year baseline, we have found the PMs for 217 stars, 150 of which are apparent cluster members. Several exhibit high PMs well beyond the escape velocity of the cluster.   However, we cannot say with confidence that we are observing any escaping stars because of probable contamination by field stars.The cluster's PM disperison is found to be 179 $\pm$ 17$  \mu$as/yr a result similar to \cite{cudworth85}.  By combining data from \cite{cudworth85} we are able to look for signs of anisotropy across the cluster's inner 150\arcsec; however we find that the distributions are consistent with isotropic orbits. Finally, the observed PM dispersion is constant with radius and we are able to put an upper limit to any central dark mass to be less than $150 d_4^3$ M$_{\odot}$ at 90\% confidence. 
\centerline{}

This work was supported by NSERC-CRSNG Canada and Gemini proposals GN-2005B-Q38, GN-2007A-Q-59 and GN-2009A-Q-29. 

~\centerline{ }



\end{document}